\documentclass[10pt,conference,compsocconf]{IEEEtran}
\usepackage{times}
\usepackage{balance}
\usepackage{caption}
\ifCLASSINFOpdf
  \usepackage[pdftex]{graphicx}
  
\else
   \usepackage[dvips]{graphicx}
\fi

\usepackage[cmex10]{amsmath}
\usepackage{algorithmic}
\usepackage[tight,footnotesize]{subfigure}

\usepackage{stfloats}

\usepackage{url}

\newcommand{\parskipsection}{\setlength{\parskip}{0.5em}}
\newcommand{\parskippresubsection}{\setlength{\parskip}{-0.4em}}
\newcommand{\parskipsubsection}{\setlength{\parskip}{0.3em}}
\newcommand{\parskipbossverse}{\setlength{\parskip}{0em}}

\usepackage{xcolor}
\usepackage{comment}
\usepackage{listings}

\DeclareCaptionFormat{listing}{\par\vskip1pt#1#2#3}
\begin{document}
\pagenumbering{gobble}
%
\title{\textbf{Secure Routine\\ \LARGE{A Routine-Based Algorithm for Drivers Identification}}\\[0.2ex]}

\author{\IEEEauthorblockN{Davide Micale\IEEEauthorrefmark{1},
Gianpiero Costantino\IEEEauthorrefmark{2}, 
Ilaria Matteucci\IEEEauthorrefmark{2}, 
Giuseppe Patan\`e\IEEEauthorrefmark{3} and
Giampaolo Bella\IEEEauthorrefmark{1}}
\IEEEauthorblockA{\IEEEauthorrefmark{1}University of Catania, Dip. di Matematica e Informatica,
Catania, Italy}
\IEEEauthorblockA{\IEEEauthorrefmark{2}Consiglio Nazionale delle Ricerche (CNR), Istituto di Informatica e Telematica (IIT), Pisa, Italy}
\IEEEauthorblockA{\IEEEauthorrefmark{3}Park Smart Srl, Catania, Italy\\\textbf{Email:} davide.micale@phd.unict.it, name.surname@iit.cnr.it, giuseppe.patane@parksmart.it, giamp@dmi.unict.it}
}


%


\maketitle

\begin{abstract}

The introduction of Information and Communication Technology (ICT) in transportation systems leads to several advantages (efficiency of transport, mobility, traffic management). However, it may bring some drawbacks in terms of increasing security challenges, also related to human behaviour.
As an example, in the last decades attempts to characterize drivers' behaviour have been mostly targeted. 
This paper presents \emph{{Secure Routine}}, a paradigm that uses driver's habits to driver identification and, in particular, to 
distinguish the vehicle's owner from other drivers. 
We evaluate Secure Routine in combination with other three existing research works based on machine learning techniques. Results are measured using well-known metrics and show that Secure Routine outperforms the compared works.
\end{abstract}


\IEEEpeerreviewmaketitle

\section{Introduction} \label{sec:introduction}
\parskipsection
Modern vehicles can be considered as computer on wheels. The mechanical parts are often controlled by software components and communication protocols are in charge of exchanging data among vehicle's components. For this reason, modern vehicles are Cyber Physical Systems \emph{(CPS)} in which used 
 technologies bring countless advantages in terms of, for instance, efficiency of city operations and services. An example among all is the Internet connectivity.
\parskipbossverse
Within this context, a problem of particular interest is how to leverage vehicular and/or smartphone data to characterize driver identification. 
Its characterization finds application in the development of software, which can be used by insurance companies to check and identify drivers or, for instance, to discourage auto theft. 
In 2019, around 56k vehicles were targeted by thieves in UK~\cite{most_stolen_car_uk}.
It equates to one car stolen every 9 minutes and 45\% of thefts occurred between midnight and 6 AM. 
Having a strategy to classify the driver's behaviour may help to mitigate this trend. 

\emph{Routine based classification} is a type of classification~\cite{6331967} that aims to find actions that are frequently repeated in time.
To complete a task, people repeat sequence of actions previously saw from others or done by themselves, no matter how tough the task is~\cite{Lavie2019}. Two persons may accomplish the same task with similar actions but with little fundamental differences~\cite{Lavie2019}. Routines can describe how people organize their lives: daily commute, weekly, meetings, holidays. Routines can also describe how a driver approaches to an intersection \cite{10.1145/2858036.2858557}. 

Based on these aspects of routine, here we introduce the paradigm of Secure Routine \emph{(SR)} that takes into account not only what the user does but also how much frequently. We use the SR paradigm within the automotive context with the aim to classify drivers. To achieve this, we elaborate and implement the SR algorithm that exploits sensors' car data, obtained, for instance, through the \emph{OBD-II}~\cite{obdii} diagnostic port. The SR algorithm evaluates the recorded data and, in particular, uses the timestamp to make an accurate classification of drivers. Then, SR leverage a Machine Learning (\emph{ML}) technique to establish driver's routines and to properly identify the driver.

To test the goodness of the Secure Routine algorithm, we compare it with other research works present in literature. The comparison is done on two different datasets and the results are evaluated using three metrics: \emph{Accuracy}, \emph{Precision} and \emph{Recall}.  Findings show that Secure Routine outperforms the compared works in all the tests carried out.

\emph{The paper is structured as follows}: next section presents the state of the art.  
In Section \ref{sec:machine_learning}, we introduce the background on ML techniques. In Section \ref{sec:secure_routine}, we present the Secure Routine paradigm used to identify drivers. Then, in Section \ref{sec:results}, we compare Secure Routine with other research works presented in literature. Finally, Section \ref{sec:conclusion} draws the conclusion of this paper and presents some hits for future research directions.

\section{State of the Art} \label{sec:related_work}
\parskipsection
In literature, there are several solutions based on ML techniques for the identification of driver's behaviour. Bernardi et al.~\cite{sijdijvj} used a Multi Layer Perception (MLP) to identify drivers. 
They used three datasets obtaining respectively 94\%, 95\% and 92\% of Accuracy. 
In particular, these results were obtained using a Start\&Stop sliding window. A sliding window combines several consecutive instances in a single instance. In particular, Start\&Stop joins instances starting when the car is moving until the car stops.
\parskipbossverse

Gao et al.~\cite{8695606} discriminated drivers through Stop-and-Go events using a \emph{voting strategy}. 
A Stop-and-Go event occurs when the car slowdowns until stops (stop phase), it stands still for five or more seconds and then speeds up (go phase). 

Wang et al.~\cite{Driver_Identification_Using_Vehicle_Telematics_Data}
identified 30 drivers by using the voting strategy and Random Forest algorithm. Authors split data and tests into different window sizes. They use six sensor signals and three derived sensor's signals along with five statistical features. With 5 minutes of testing data this model achieves almost 93\% of Accuracy. With a sliding window of 5 seconds and 6 minutes of testing data they achieve 100\% of Accuracy.

Girma et al. in ~\cite{muflo} used the Long Short-Term Memory (LSTM) algorithm with sliding windows and tested their model on \cite{korean_dataset} and \cite{brazilian_dataset} datasets with Precision and Recall of 98\%.

Kwak et al. in~\cite{kwak2016know} selected 15 features to identify drivers behaviour. For each feature they computed the mean, median and standard deviation according to a reference sliding window. Thus, the total number of features is 45. They used different ML algorithms and achieved the best Accuracy of 99,6\% applying Random Forest on \cite{korean_dataset} dataset.

Martinelli et al. in~\cite{cnr_driver_recognition2} tested several Decision Tree algorithms with the same dataset \cite{korean_dataset} using all 51 features. They obtained a Precision and Recall equal to 99,2\% with J48. The same authors in \cite{cnr_driver_recognition1} used only six features out of 51 features of \cite{korean_dataset} dataset. In this case, Precision and Recall decreased to 98,9\% due to under-fitting. 

Compared to our paper, \cite{sijdijvj}, \cite{8695606}, \cite{Driver_Identification_Using_Vehicle_Telematics_Data}, \cite{kwak2016know}, \cite{cnr_driver_recognition2} and \cite{cnr_driver_recognition1} do not look for frequency. Also, LSTM in \cite{muflo} obtained lower scores in comparison with a \emph{Decision Tree} \emph{(DT)} algorithm (\cite{kwak2016know}, \cite{cnr_driver_recognition2} and \cite{cnr_driver_recognition1}) on the same dataset. As shown by \cite{cnr_driver_recognition1}, certain features discriminate better than others for some drivers. Hence, SR must use the best feature set for each driver. \cite{sijdijvj}, \cite{cnr_driver_recognition1} and \cite{cnr_driver_recognition2} are the only ones that make owner-driver identification, they select the same feature set for all drivers. Finally, SR breaks down the timestamp in fine grained units to detect frequency in order to increase the accuracy. 

\section{Machine Learning} \label{sec:machine_learning}
\parskipsection
ML is the study of computer algorithms that improve automatically through experience. 
ML algorithms build a mathematical model to make predictions or decisions without being explicitly programmed to do so.
At the basis of the model, there is a dataset that has to be processed.
Such dataset can be considered as a table in which all data are listed. 
Each row of the table is called \emph{instance} and each column represents a \emph{feature} of the instance.
The dataset is usually split into two parts, the \emph{training dataset} and the \emph{test dataset}.
The model is created on the basis of the training dataset. 
Instead, a \emph{test dataset} represents all instances adopted to verify how much accurate our model is in doing the classification. 
\parskipbossverse

ML techniques are largely adopted for the identification and classification of users.
In the following, we introduce an example of ML algorithm based on DT predictive modelling approach.
A DT consists on 
a tree data structure that contains rules 
to classify the instance.
For each level of the tree, the value of a feature of the instance is tested, for example, through a specific question. Each internal node of the tree contains a test. Depending on the answer, the model follows a different edge: the left edge if the result of the test is true,  otherwise the right edge is followed. Finally, the leaf nodes, i.e., the nodes with no children, contain the prediction.

\parskippresubsection
\subsection{Decision Tree Requirements}
\parskipsubsection

A DT algorithm must create a tree with the minimum number of levels. 
This allows the ML algorithm to classify the instance as fast as possible.
To build a DT with a low number of levels, it is necessary to select the best tests for the model. 
This is done by selecting the appropriate \emph{Formula} to make the selection. 
A \emph{Formula} specifies the criterion chosen to establish which is the next test to perform in the DT. 

\parskipbossverse

\begin{figure*}[t]
	\centerline{
		\subfigure[DT with two levels.]{\label{fig:sub1}\includegraphics[width=2.5in]{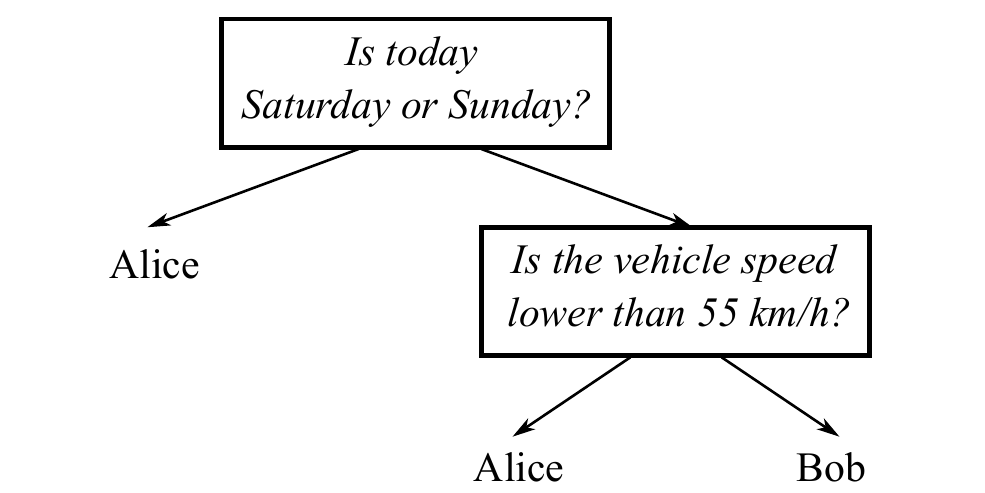}}
		\hfill
		\subfigure[DT with one level.]{\label{fig:sub2}\includegraphics[width=2.5in]{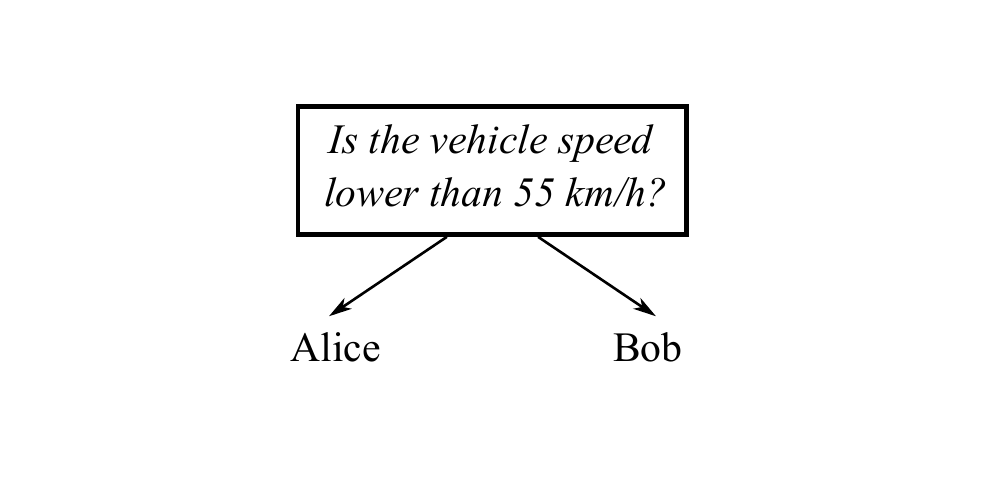}}
	}    
	
	\caption{Comparison of two possible DT for solving the same problem.}
	\label{fig:test}
\end{figure*}

For instance, let 
\emph{Alice} and \emph{Bob} be two drivers that are used to going on the Sixth Avenue. Alice goes on the Sixth Avenue all days of the week, instead Bob goes only from Monday to Friday. Bob drives slightly faster than Alice, with a speed up to 55 Km/h. 
A possible DT model is the one in Figure~\ref{fig:sub1} that is built by putting on the tree root the following test:
\begin{center}
	\lq\lq \emph{Is today Saturday or Sunday?}\rq\rq
\end{center}
Following the root test, we have that the left child is taken by Alice instead the right child corresponds to the following test:
\begin{center}
	\lq\lq \emph{Is the vehicle speed lower than 55 Km/h?}\rq\rq
\end{center}
Again the left child is a leaf node that represents Alice, whereas the right child is the leaf node representing Bob.

Despite the above DT model is a valid model for our example, we may produce a better tree in which a root node is configured with the following test (Figure~\ref{fig:sub2}):
\begin{center}
	\lq\lq \emph{Is the vehicle speed lower than 55 Km/h?}\rq\rq
\end{center}

In this case, the left child is the leaf node Alice and the right child is the leaf node Bob. Hence, a ML algorithm concludes its prediction with only one test.

A DT has to be simple. 
This allows the DT to be flexible enough to represent also further instances. 

Thus, if the built model is too complex, it may not represent new labelled instances, i.e., for instance those ones present in a test-set. This may cause a high error rates, generating the \emph{over-fitting} error. To reduce the over-fitting error, the \emph{pruning} technique can be adopted to obtain a simpler version of the tree by pruning some nodes.
Another solution to mitigate the over-fitting error is the \emph{feature selection} that works by removing features. However, pruning too many nodes and removing too many features or relevant ones may lead to higher error rates, aka \emph{under-fitting}.

\parskippresubsection
\subsection{Decision Tree Algorithms}
\setlength{\parskip}{0.5em}
Several DT algorithms were developed to generate models.
\parskipbossverse
The \emph{C4.5} was proposed in 1993~\cite{Quinlan1993} and it uses the Gain Ratio (GR) of a feature ``{}X" of the training set (T) to establish which is the best test to perform. 
\small{
	\begin{equation}
		GR = { H(T) - H(T|X) \over{H(X)} }
	\end{equation}
}
\normalsize
where:
\begin{itemize}
	\item $H(T)$ indicates the \emph{entropy} of T, i.e., the quantity of information carried by the probability distribution of labels in T~\cite{shannon_entropy}, calculated as:
	\small{
		\begin{equation}
			H(T) = - \displaystyle\sum_{j=1}^{k} {freq(C_j, T) \over{|T|}} \times log_2 \left({freq(C_j, T) \over{|T|}} \right)
		\end{equation}
	}
	\normalsize
	where:
	\begin{itemize}
		\item $k$ is the number of classes;
		\item $freq(C_j, T)$ is the number of instances in the j-th class; 
		\item $|T|$ is the number of instances of T.
	\end{itemize} 
	\item $H(T|X)$ indicates the entropy after partitioning T in ``{}n" parts, where ``{}n" is the number of possible values assumed by X: 
	\small{
		\begin{equation}
			H(T|X) = \sum_{i=1}^{n} {|T_i| \over{|T|}} \times H(T_i)
		\end{equation}
	}
	\normalsize
	where:
	\begin{itemize}
		\item $|T_i|$ is the number of instances with the i-th value assumed by the feature X;
		\item $ H(T_i)$ indicates the entropy of the set of instances with the i-th value assumed by the feature X.
	\end{itemize} 
	\item $H(X)$ indicates the entropy of X:
	\small{
		\begin{equation}
			H(X) = - \displaystyle\sum_{i=1}^{n} {|T_i| \over{|T|}} \times log_2 \left({|T_i| \over{|T|}} \right)
		\end{equation}
	}
	\normalsize
\end{itemize}

Note that C4.5 can handle features with unknown values and real numbers and may make use of the pruning technique.
 
\emph{Random Forest} (RF)~\cite{random_forests}\cite{Breiman2001} is an algorithm formed by a set of DTs. Each tree is built from a random sampling with replacement of the training-set. 
Each node of a tree is the best test defined on a subset of features, instead of on all available ones. Trees are not pruned. In prediction phase, an instance is run on each tree and each tree makes a prediction. The most predicted value becomes the prediction of RF. Also, RF includes a procedure in case of unknown values in the dataset.

\section{Secure Routine} \label{sec:secure_routine}
\parskipsection
In literature, the concept of \emph{Routine} is already exploited to classify users or drivers~\cite{10.1145/2858036.2858557}.
A Routine is defined as a set of actions that a person frequently perform in response to a circumstance \cite{10.1093/oxfordjournals.cje.a013692}. Hence, routines can describe how people organize their lives: daily commute, weekly, meetings, holidays.
Here, we refine the concept of Routine by introducing the paradigm of \emph{Secure Routine} that takes into account not only what the user does but also how much frequently.
\parskipbossverse

We define SR and present its application into the automotive context to perform driver's behavioural identification. 
To this aim, SR analyses all tracking data recorded by vehicle's sensors while the user is driving it. 
Tracked data are organized in separate instances according to the sensor that collects them and the timestamps when the event occurs.
Hence, SR firstly decomposes the timestamp of each instance and extracts second, minute, hour, day of week, day, month and year. Then, SR removes less relevant features, as we will describe below using the Feature Selection~\emph{(FS)} technique. Successively, the data collected by sensors are correlated with the timestamp previously decomposed. Then, a ML algorithm examines these data. 
The output is a model representing users' Secure Routine. As final step, the obtained model is compared with an observed user's driver behaviour for his/her identification.

To show the value added by the Secure Routine to identify drivers, we introduce the following example. Let us consider Alice and Bob who are used to going on the Sixth Avenue. Alice usually goes there at 12PM, and Bob at 7PM. If we do not consider the timestamp information, the resulting model of Alice and Bob will contain only the information \emph{``{}The user is used to going on the Sixth Avenue"}. 
In this situation, the observed behaviour will be compared to understand whether the driver is Alice or Bob. However, this selection is quite difficult since the missing timestamp information is fundamental to distinguish between the drivers. On the contrary, if we consider also the timestamp in which the event happens, the identification will be unique in this case. In fact, if the vehicle is at 7PM on the Sixth Avenue, therefore the driver is Bob. 

This is what Secure Routine does considering daily routines as well as monthly and yearly ones. 
Hence, SR may be very useful, for instance, to mitigate scenarios as the one depicted in Section~\ref{sec:introduction}: in UK cars are often stolen at night. If the vehicle's owner does not usually drive during the night, SR can easily detect the weird behaviour. 
In particular, the SR paradigm is built upon a ML algorithm that uses as training-set the data recorded through an OBD-II device. A closer working mechanism of SR is presented in~\cite{10.1007/978-3-642-18178-8_9}. Here, the authors prefer to involve the interval between a rerun of the same action. Let us consider this other example in which Alice goes on the Sixth Avenue every 24 hours for the whole week, instead Bob every 24 hours from Monday to Friday. In this case, the routine of Bob will be modelled as intervals of 24 and 72 hours. So, if we consider a driver moving on Saturday, we would not be able to identify the driver, neither Alice nor Bob, since the interval is set to 24 hours. On the contrary, if the day of the week is taken into account, Alice will be correctly identified.

\parskippresubsection
\subsection{SR Algorithm}
\parskipsubsection
Let us consider a target vehicle belonging to a driver $d$. 
The SR algorithm acts in four phases: 
\parskipbossverse
\paragraph{Model Generation Dataset}
Whenever a vehicle is used, its sensors register pieces of information about several \emph{features}, e.g., the water temperature, the speed, the brake pressure, and so on. We assume to take trace of all these data in combination with the timestamp in which each instance of data is generated. Data are taken from the OBD-II port by using an OBD-II interface~\cite{olm327}. Each instance of data is called \emph{interaction} of the driver $d$ with the vehicle and it is denoted as $in _{i, d}$ where $i$ is the timestamp. Interactions are composed by the timestamp, recorded with the following template:  
(day, month, year, hour, minute, second and day of the week) plus the others features obtained from the OBD-II.

\paragraph{FS paradigm} To mitigate the possible over-fitting error, we implement the \emph{FSParadigm} (Figure~\ref{code:FSParadigm}). 

\begin{figure}[b]
	\small{ \begin{lstlisting}[language=C, numberstyle=\tiny,  stepnumber=1]
        function FSParadigm(instances)
            ranking $\leftarrow$ GR(instances)
            ranking$_{ordered} \leftarrow$ order ranking ascending
            features$_{>0} \leftarrow$ discard features with rank = 0 from ranking$_{ordered}$
            (features$_{no\_timestamp\_correlated}$, features$_{timestamp\_correlated}) \leftarrow$ features$_{>0}$
            ranking$_{no\_timestamp\_correlated} \leftarrow$ ranking from ranking$_{ordered}$ of features present in features$_{no\_timestamp\_correlated}$
            average$_{ranking} \leftarrow$ mean(ranking$_{no\_timestamp\_correlated})$
            subset$_{no\_timestamp\_correlated} \leftarrow $ discard features sum is less than or equal to the average$_{ranking}$ from ranking$_{no\_timestamp\_correlated}$
            subset $\leftarrow$ subset$_{no\_timestamp\_correlated}$ $\cup$ features$_{timestamp\_correlated}$
            return subset
        \end{lstlisting}
        }
    \caption{Feature Selection Paradigm}
    \label{code:FSParadigm}
\end{figure}

\normalsize

\emph{FSParadigm} is designed to select the best features to use. It firstly ranks all features applying the \emph{Gain Ratio} approach and then features are sorted in ascending order. Those features with rank equal to zero are discarded. Then, the average-rank among all features not correlated to the timestamp is calculated. The FS discards those features, except those related to time, whose rank sum is less than or equal to the average-rank.

\paragraph{Model Generation Algorithm} Let us consider that a vehicle may be driven by $d$ but also by other people, e.g., friends or relatives of $d$.  
In the modelling phase, our algorithm (Figure~\ref{code:SRalgorithm}) considers all the past interactions recorded by the vehicle and labels with 1 each interaction that belongs to $d$, 0 otherwise. 
The labelled interactions are sent to a DT algorithm that generates the model for the driver $d$. 

\begin{figure}[b]
	\small{ \begin{lstlisting}[language=C, numberstyle=\tiny,  stepnumber=1]
        function generate_model(d)
            ins$_d$ $\leftarrow$ get interactions from $db$ made by $d$, labeling $1$
            ins$_o$ $\leftarrow$ get interactions from $db$ made by others, labeling $0$
            ins$_{all}$ $\leftarrow$ ins$_d$ $\cup$ ins$_o$
            subset $\leftarrow$ FSParadigm(ins$_{all}$)   
            model $\leftarrow$ MLAlgorithm(ins$_{all}$ with features from subset)
            return model 	 
         \end{lstlisting}
        }
    \caption{Secure Routine Model Generation}
    \label{code:SRalgorithm}
\end{figure}

\normalsize
In particular, in line 5, \emph{FSParadigm} is the Feature Selection paradigm we described above as part of SR and line 6 (\emph{MLAlgorithm}) indicates the ML algorithm in use with the subset of features obtained before.

\paragraph{SR Identification strategy}

Once the model is generated, SR makes the identification evaluating each interaction. In particular, SR links an interaction to the vehicles' owner if the ML algorithm predicts and labels it as $1$, otherwise $0$.

\section{Secure Routine Evaluation} \label{sec:results}
\parskipsection
We evaluated Secure Routine in two steps: first, we run it using two ML algorithms and we verified which of them best performs to identify drivers.
Then, we compared Secure Routine with the following research works present in literature:
\parskipbossverse
\begin{itemize}
 \item 	Martinelli et al.~\cite{cnr_driver_recognition1} referred in the following as $M$.
 \item Kwak et al.~\cite{kwak2016know} referred in the following as $K$.
 \item Girma et al.~\cite{muflo} referred in the following as $G$.
\end{itemize} 
\subsection{Datasets}

We run the experiments using two datasets presented in~\cite{korean_dataset}, referred as $\Theta$, and~\cite{best}, referred as $\Psi$. The former is a dataset used also by $M$, $K$ and $G$ in their research works. So we can fairly make a comparison. However, the $\Theta$ dataset does not contain a fundamental feature used by SR, this is the \emph{timestamp} of each represented instance. Nevertheless, $\Theta$ dataset contains the \emph{engine runtime} that provides the minutes to be used as timestamp needed for SR to work.

On the other hand, $\Psi$ dataset contains a timestamp for each instance by default. This feature allows Secure Routine to fully work by using all available pieces of information. In particular, SR expands the timestamp to generate all time dependent features. As far as we know, the other compared research works do not make use of this dataset to evaluate their proposal. So, to evaluate SR even in this case, we were able to re-run the work proposed by Martinelli et al. and calculate the results for the owner-driver identification. On the other side, the works $K$ and $G$ did not calculate the owner-driver identification and, also, it was not possible to re-run their algorithms since the implementation is not publicly available. In the specific case of $G$, the authors published only the pre-built model and we were not able to use it.

\parskippresubsection
\subsection{Metrics}\label{results:metrics}
\parskipsubsection
To get a comparable result of SR with $M$, $K$ and $G$, we evaluate \emph{Accuracy}~\cite{FIRST_CONTACT_WITH_DEEP_LEARNING_PRACTICAL_INTRODUCTION_WITH_KERAS}, 
\emph{Precision} and \emph{Recall}~\cite{cnr_driver_recognition1}. 
\parskipbossverse
\begin{itemize}
    \item \emph{Accuracy} represents how often the model is making a correct prediction. It is the ratio between the number of correct predictions and the number of predictions: 
    \small{
        \begin{equation}
            Accuracy = { TP + TN \over{TP + TN + FP + FN} }
        \end{equation}
    }
    \normalsize
    where:
    \begin{itemize} 
        \item TP (True Positive) is the number of instances belonging to the vehicle's owner that are correctly predicted;
        \item TN (True Negative) is the number of instances not belonging to the vehicle's owner that are correctly predicted; 
        \item FP (False Positive) is the number of instances belonging to another person but incorrectly predicted;
        \item FN (False Negative) is the number of instances belonging to the vehicle's owner but incorrectly predicted.
    \end{itemize}

    \item \emph{Precision} measures how often the predicted instances belonging to the vehicle's owner are true. It is calculated as the ratio between $TP$ and $TP + FP$:
    \small{
        \begin{equation}
            Precision = { TP \over{TP + FP} }
        \end{equation}
    }
    \normalsize
    
    \item \emph{Recall} identifies how often the instances belonging to vehicle's owner are correctly predicted. It is calculated as the ratio between $TP$ and $TP + FN$:
    \small{
        \begin{equation}
            Recall = { TP \over{TP + FN} }
        \end{equation}
    }
    \normalsize
    
\end{itemize}

To better estimate the three metrics depicted above, in our experiments we used the 10-fold cross-validation~\cite{A_Study_of_Cross-Validation_and_Bootstrap_for_Accuracy_Estimation_and_Model_Selection} approach. First, we split the dataset on 10 equal size subsets $D_1$, $D_2$, ..., $D_{10}$. Each instance of the dataset is randomly inserted in a subset. 
Then, we constructed 10 training sets $Tr_1$, $Tr_2$, ..., $Tr_{10}$ and 10 testing sets $Te_1$, ..., $Te_{10}$. $Tr_i$ is made of all subsets except $D_i$ and $Te_i$ is made of $D_i$ with $i \in \{1,2, ..., 10\}$. For each pair $(Tr_i, Te_i)$ is calculated $Accuracy_i$, $Precision_i$ and $Recall_i$. Finally, we calculated the final value of $Accuracy$, $Precision$ and $Recall$ as the mean of $Accuracy_i$, $Precision_i$ and $Recall_i$, respectively. 

\begin{figure*}[b]
    \centerline{
        \subfigure[Number of instances for each driver in $\Theta$]{\label{fig:instances_vehicles_kym}\includegraphics[width=2.5in]{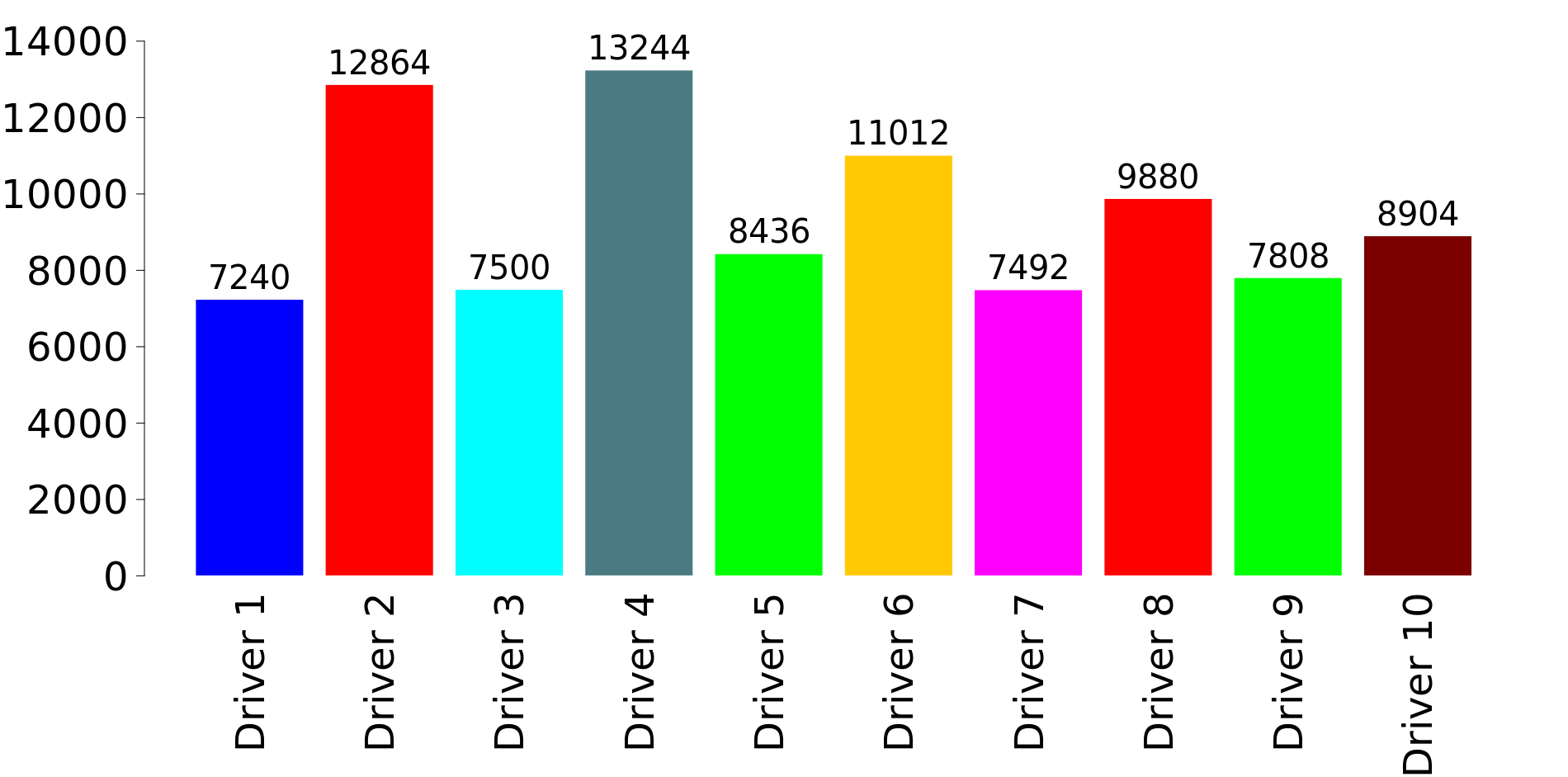}}
        \hfill
        \subfigure[Number of instances for each driver in $\Psi$]{\label{fig:instances_vehicles}\includegraphics[width=2.5in]{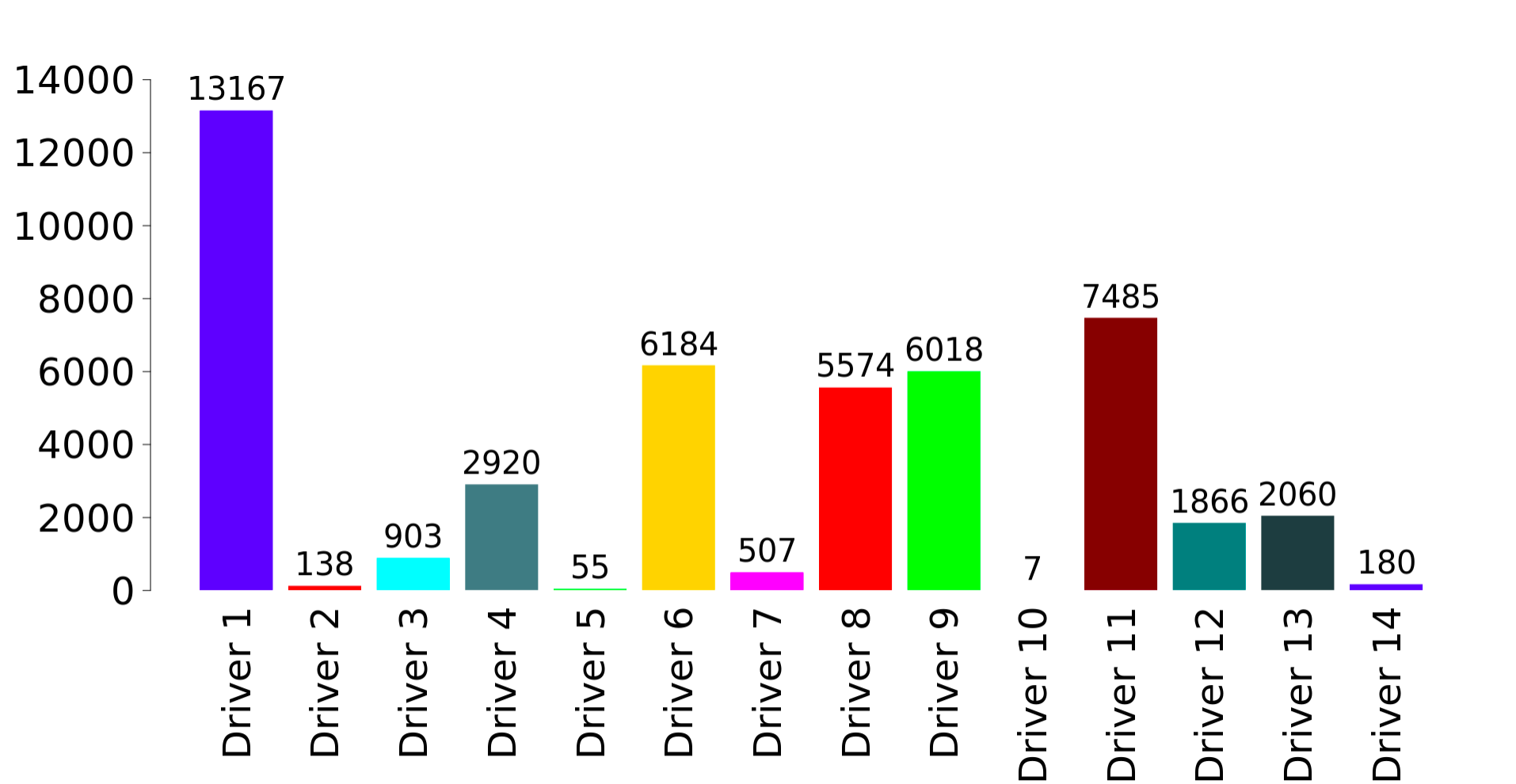}}
    }    
    
    \caption{Driver distributions on the datasets.}
    \label{fig:distributions}
\end{figure*}

\parskippresubsection
\subsection{Experiments}
\parskipsubsection
We performed four types of experiments to evaluate Secure Routine. The first experiment is related to 
multi-driver identification problem~\cite{cnr_driver_recognition1}, i.e., properly identify who is the driver. However, as step zero, we decided to find the most suitable ML algorithm with the best features set to evaluate SR. 
We leverage on Weka~\cite{una} as software that contains a collection of visualization tools and algorithms for data analysis and predictive modelling. So, we used the available Gain Ratio method to rank each feature. Then, we employed \emph{J48}, which is the implementation of the C4.5 algorithm, and RF algorithm over the driver identification.
\parskipbossverse

In this step, results are obtained on $\Theta$ dataset. It contains data from 10 drivers. Figure~\ref{fig:instances_vehicles_kym} shows the driver instances' distribution. Drivers have 9438 instances on average: Driver 4 has the highest number of instances with 13244 samples while Driver 1 has the lowest number with 7240 instances. In addition, drivers drove two times in the same path in similar time-window. Dataset instances are recorded per second.

Table~\ref{table:best_algorithm_features_exp1} shows the results obtained comparing SR implemented into J48 and RF 
algorithms applied to the driver identification using $\Theta$ dataset.
RF algorithm with feature selection (37 features) obtained the best Precision and Recall.

\captionsetup{font={footnotesize,sc},justification=centering,labelsep=period}%
\begin{table*}[hb] 
    \caption{Comparing SR using J48 and Random Forest over the multi-driver identification problem.}
	\centering
	\begin{tabular}{|c|l|c|l|c|l|c|l|c|l|c|l|c|l|}
		\hline
		\multicolumn{4}{|c|}{\textit{J48}}                                                                                					  & \multicolumn{4}{c|}{Random Forest}                                                                                                                                                                                                                               \\ \hline
		\multicolumn{2}{|c|}{All features}                         & \multicolumn{2}{c|}{Feature selection} 					 	  & \multicolumn{2}{c|}{All features}                         & \multicolumn{2}{c|}{Feature selection} 							\\ \hline
		Precision                    & \multicolumn{1}{c|}{Recall} & Precision                       & \multicolumn{1}{c|}{Recall}    & Precision                   & \multicolumn{1}{c|}{Recall} & Precision                      & \multicolumn{1}{c|}{Recall}    \\ \hline
		\multicolumn{1}{|l|}{99,2\%} & 99,2\%                      & \multicolumn{1}{l|}{99,3\%}     & 99,3\%                         & \multicolumn{1}{l|}{99,3\%} & 99,3\%                      & \multicolumn{1}{l|}{99,6\%}    & 99,6\%                         \\ \hline
	\end{tabular}
    \label{table:best_algorithm_features_exp1}
\end{table*}
\captionsetup{font={footnotesize,rm},justification=centering,labelsep=period}%

After selecting SR with RF and the most appropriate features ranked by the Gain Ratio method, we show the first experiment results obtained by comparing SR with the work in $M$, $K$ and $G$ on the $\Theta$ dataset.
As shown in Table \ref{table:comparison}(a), Secure Routine and $K$ achieves the best results. Note that $M$ did not calculate the accuracy in the paper, so we established this value through the replication of their experiment. Instead, $K$ did not provide on their research Precision and Recall. Finally, for $G$ we were not able to retrieve the exact Accuracy.

As we can see in Table \ref{table:comparison}(b), SR achieves almost a perfect Precision, i.e., 100\%, but with the worst Recall and this depends on the features selection. In fact, if we increment the number of features, we increase the Recall but the Precision is decreased. Here, we decided to obtain a higher Precision selecting the most appropriate features using the Gain Ratio.

\captionsetup{font={footnotesize,sc},justification=centering,labelsep=period}%
\begin{table*}[t]
    \caption{Comparison of Secure Routine with related works}
    \centering{
        \subtable[Comparison of Secure Routine with $M$, $K$ and $G$.]{
            \begin{tabular}{|c|l|c|l|c|l|c|l|}
                \hline
                \multicolumn{2}{|c|}{Secure Routine}                       & \multicolumn{2}{c|}{$M$}                    			   & \multicolumn{2}{c|}{$K$}                      	   & \multicolumn{2}{c|}{$G$}                                                  			\\ \hline \hline
                Precision                    & \multicolumn{1}{c|}{Recall} & Precision                   & \multicolumn{1}{c|}{Recall} & Precision               & \multicolumn{1}{c|}{Recall} & Precision                                          & \multicolumn{1}{c|}{Recall}   \\ \hline
                \multicolumn{1}{|l|}{99,6\%} & 99,6\%                      & \multicolumn{1}{l|}{99,2\%} & 99,2\%                      & \multicolumn{1}{l|}{N.A.} & N.A.                          & \multicolumn{1}{l|}{{\color[HTML]{000000} 98,8\%}} & {\color[HTML]{000000} 98,1\%} \\ \hline \hline
                \multicolumn{2}{|c|}{Accuracy}                             & \multicolumn{2}{c|}{Accuracy}                             & \multicolumn{2}{c|}{Accuracy}                         & \multicolumn{2}{c|}{Accuracy}                                                      \\ \hline
                \multicolumn{2}{|l|}{99,6\%}                               & \multicolumn{2}{l|}{99,2\%}                               & \multicolumn{2}{l|}{99,6\%}                           & \multicolumn{2}{l|}{N.A.}                                                            \\ \hline
            \end{tabular}
            \label{table:comparison_exp1}
        }
        \hfill
        \subtable[Comparison of Secure Routine with $M$.]{
            \begin{tabular}{|c|l|c|l|}
                \hline
                    \multicolumn{2}{|c|}{Secure Routine} 											 & \multicolumn{2}{c|}{$M$}                                               						  \\ \hline \hline
                    Avg. Precision                      & \multicolumn{1}{c|}{Avg. Recall}        	 & Avg. Precision                                           & \multicolumn{1}{c|}{Avg. Recall}    \\ \hline
                    \multicolumn{1}{|c|}{99,8\%}        & 98,5\%                             		 & \multicolumn{1}{l|}{{\color[HTML]{000000} 99,3\%}} & {\color[HTML]{000000} 99,3\%} 			  \\ \hline
            \end{tabular}
            \label{table:comparison_exp2}
        }      
    }
    \centering{
        \subtable[Comparison of Secure Routine with $M$ for multi-driver identification.]{
            \begin{tabular}{|c|l|c|l|c|l|}
                \hline
                \multicolumn{2}{|c|}{Secure Routine} 								 & \multicolumn{2}{c|}{$M$}                    				 \\ \hline \hline
                Precision                         & \multicolumn{1}{c|}{Recall}      & Precision                   & \multicolumn{1}{c|}{Recall} \\ \hline
                \multicolumn{1}{|c|}{99,4\%}      & 99,4\%                           & \multicolumn{1}{l|}{90,4\%} & 89,8\%                      \\ \hline
            \end{tabular}
            \label{table:comparison_exp3}
        }
        \hfill
        \subtable[Comparison of Secure Routine with $M$ for owner identification.]{
            \begin{tabular}{|c|l|c|l|c|l|}
                \hline
                \multicolumn{2}{|c|}{Secure Routine} 										   & \multicolumn{2}{c|}{$M$}                    						 \\ \hline \hline
                Avg. Precision                         & \multicolumn{1}{c|}{Avg. Recall}      & Avg. Precision                   & \multicolumn{1}{c|}{Avg. Recall} \\ \hline
                \multicolumn{1}{|c|}{99,6\%}       	   & 98,1\%                           	   & \multicolumn{1}{l|}{95,1\%} 	  	  & 82,9\%                      	 \\ \hline
            \end{tabular}
            \label{table:comparison_exp4}
        }
    }      
    \label{table:comparison}
\end{table*}
\captionsetup{font={footnotesize,rm},justification=centering,labelsep=period}%
The second experiment is related to the \emph{Owner Driver identification}, i.e., does the instance belong to the vehicle's owner? In this case, we compared SR only with $M$ since $K$ and $G$ did not calculate the owner driver identification. As stated by the authors of $M$, they use the same feature set for both the multi-driver and owner driver identification.

The third experiment that we propose is related to the multi-driver identification on the $\Psi$ dataset. This contains data from 14 drivers. Figure~\ref{fig:instances_vehicles} shows that drivers' instances are not equally distributed. For example, Driver 1 has the highest number of instances with 13617 samples, whereas Driver 10 has the lowest number with only 7 instances. This may depend on the fact that some users drive frequently whereas other users rarely. However, 7 instances are not enough to build a model for the Driver 10. So, we decided to exclude Driver 10 instances in our experiments to not alter the final result. Also, many instances contain empty values because of errors on gathering data. Instances are recorded every 7 seconds.

Compared to the $\Theta$ dataset, $\Psi$ contains by default 32 features. Nevertheless, five of these features are timestamp related and are \emph{minute}, \emph{hour}, \emph{day of the week}, \emph{month}, \emph{year}. Other features, such as, \emph{model}, \emph{car\_year}, are removed since they do not give any useful information about the user driving style. The dataset also contains \emph{engine\_runtime} from which we extract \emph{engine\_runtime\_minute}.

In this experiment, we used the GR method for features selection. Starting from pruned $\Psi$ dataset, we evaluated SR. As previously stated, we know that there are no other research works that use this dataset. So, we had only the possibility to replicate the best solution proposed by $M$.

Table~\ref{table:comparison}(c) shows that Secure Routine with feature selection achieves the best result both for Precision and Recall.

To conclude the evaluation, last experiment focused on the owner driver identification. Table \ref{table:comparison}(d) indicates that SR with features selection has the best performance when compared with $M$. SR obtained an average precision of 99,6\%, which means that for 8 drivers SR established a perfect Precision whereas $M$ achieved this Precision only for 4 drivers with an average Precision of 95,1\%. Regarding the Recall, SR largely outperformed $M$ in percentage and  SR achieved a perfect Recall score for one driver, whereas $M$ never obtained a perfect Recall.

\section{Conclusion and future work} \label{sec:conclusion}
\parskipsection
In this paper, we introduced for the first time the Secure Routine paradigm 
\parskipbossverse
to identify the vehicle's owner taking into account the driving style. Also, we presented the algorithm implemented by means of machine learning algorithms and we showed how SR works to identify the driver. Then, we compared SR with other three existing research papers and we evaluated them considering Precision, Accuracy and Recall metrics. Experiments made use of two different datasets. Findings showed that SR obtains the best results compared with the other algorithms considering both experiments regarding the identification of the vehicle's owner and the multi-driver.

As future work, we plan to improve the algorithm of Secure Routine by considering additional features to increase its identification capabilities, i.e., statistical features. We will also improve our \emph{FSParadigm} to enable a better feature selection. 

\section*{Acknowledgment} This work has been partially supported by the COSCA research project (NGI\_TRUST 2nd Open Call 2019002).

\balance
\bibliographystyle{IEEEtran}
\bibliography{bibtex/global.bib}

\end{document}